# Endohedral Confinement of a DNA Dodecamer onto Pristine Carbon Nanotubes and the Stability of the Canonical B Form

Fernando J.A.L. Cruz,[1,2,*] Juan J. de Pablo[2,3] and José P.B. Mota[1]

[1]Requimte/CQFB, Departamento de Química, Faculdade de Ciências e Tecnologia, Universidade Nova de Lisboa, Caparica, 2829-516, Portugal. E-mail: fj.cruz@fct.unl.pt
[2]Department of Chemical and Biological Engineering, University of Wisconsin-Madison, Madison, Wisconsin, 53706, USA
[3]Institute of Molecular Engineering, University of Chicago, Chicago, Illinois, 60637, USA

Although carbon nanotubes are potential candidates for DNA encapsulation and subsequent delivery of biological payloads to living cells, the thermodynamical spontaneity of DNA encapsulation under physiological conditions is still a matter of debate. Using enhanced sampling techniques, we show for the first time that, given a sufficiently large carbon nanotube, the confinement of a double-stranded DNA segment, 5'-D(*CP*GP*CP*GP*AP*AP*TP*TP*CP*GP*CP*G)-3', is thermodynamically favourable under physiological environments (134 mM, 310 K, 1 bar), leading to DNA-nanotube hybrids with lower free energy than the unconfined biomolecule. A diameter threshold of 3 nm is established below which encapsulation is inhibited. The confined DNA segment maintains its translational mobility and exhibits the main geometrical features of the canonical B form. To accommodate itself within the nanopore, the DNA's end-to-end length increases from 3.85 nm up to approximately 4.1 nm, due to a ~ 0.3 nm elastic expansion of the strand *termini*. The canonical Watson-Crick H-bond network is essentially conserved throughout encapsulation, showing that the contact between the DNA segment and the hydrophobic carbon walls results in minor rearrangements of the nucleotides H-bonding. The results obtained here are paramount to the usage of carbon nanotubes as encapsulation media for next generation drug delivery technologies.

## I. INTRODUCTION

Deoxyribonucleic acid (DNA) and single-walled carbon nanotubes (SWCNTs) are prototypical one-dimensional structures; the former plays a central role in chemical biology and the latter holds promise for nanotechnology applications [1-3]. From the point of view of biological purposes and DNA manipulation, carbon nanotubes have been proposed to be used as templates for DNA encapsulation [4, 5], intracellular penetration via endocytosis and delivery of biological payloads [6, 7], and ultrafast nucleotide sequencing [2, 8, 9]. While structure in its natural form and environment is well established (*e.g.* B-DNA in aqueous solution), their interactions have been the subject of intense investigation [4, 5, 8, 10-16], nonetheless, the corresponding molecular-level phenomena remain rather unexplored. Is confinement thermodynamically spontaneous (free-energy)? How important are the conformational properties of the encapsulated double-strand (entropy)? How does the confinement process depend on nanotube diameter? Moreover, previous work has focused almost exclusively on exoadsorption of DNA on the external surface of SWCNTs [8, 11, 17-19], overlooking the possibility of endohedral confinement. Nonetheless, it is well known that the conformational properties of biopolymers under confinement are of crucial relevance in living systems (*e.g.*, DNA packing in eukaryotic chromosomes, viral capsids).

DNA encapsulation is a phenomenon that remains utterly unmapped. Most of the earlier work focused on temperatures markedly distinct from the physiological value, precluding extrapolation of results to *in vivo* conditions. The pioneer work of Lau *et al.* [5] showed that a small DNA strand, initially confined in a 4 nm diameter nanotube, exhibits dynamics similar to the unconfined molecule, but that behaviour is drastically altered when diameter is decreased to 3 nm. Because the DNA had been artificially inserted, no information about the encapsulation process itself, *e.g.*, its thermodynamical spontaneity and kinetics, was provided. Previous experiments and calculations indicate that the biomolecule can be confined onto $D$ = 2.7 nm SWCNTs [13] and $D$ = 3–4 nm multi-walled carbon nanotubes (MWCNTs) [20, 21], however, the corresponding working temperatures (350 – 400K) were far too high to have any physiological relevance. Furthermore, the experimental observations of Iijima *et al.* indicated that encapsulation of DNA onto MWCNTs was a competing mechanism with wrapping of the biomolecule around the external wall [20]; their reported data failed to identify the relevant conditions upon which the confinement process is favoured, such as ionic strength of the media and temperature. Recently, Mogurampelly and Maiti addressed [22] the encapsulation of dsDNA and siRNA onto SWCNTs, and established a critical diameter of $D$ = 2.67 nm and $D$ = 2.4 nm for the former and the latter, respectively, bellow which confinement was completely inhibited, and attributed it to a large free-energy barrier associated with the nanopore entrance.

Our results obtained under precise physiological conditions (310 K, 1 bar, [NaCl] = 134 mM) show that an atomically detailed DNA dodecamer can be encapsulated onto a $D$ = 4 nm SWCNT, resulting in a decrease of the system's free energy. Encapsulation kinetics is fast (< 16 ns) and the double strand retains its translational mobility within the nanotube. Very interestingly, our data indicate that the encapsulated molecule free-energy minima correspond to a DNA conformation similar to the bulk canonical B form [23], with a pitch length of 3.4 nm (10 bps repeating unit) and a double strand end-to-end length of approximately 4.1 nm. The canonical Watson-Crick H-bonds network is roughly maintained throughout confinement, exhibiting probability distributions essentially corresponding to more than 75 % of hydrogen bonds existence. As far as we are aware these observations are the first of their kind, and they come to pave the way for the design of smart nanotube based devices for *in vivo* DNA encapsulation.

## II. MODELS AND METHODS
### A. Molecular Models

All molecules in this work are described using atomistically detailed force fields, including electrostatic charges in each atom. The dispersive interactions are calculated with the Lennard-Jones (12,6) potential, cross parameters between unlike particles determined by the classical Lorentz-Berthelot mixing rules, and electrostatic energies described by Coulomb's law. DNA is modelled as a completely flexible molecule within the framework of the AMBER99sb-ildn force-field [24, 25], the corresponding potential energies associated with bond stretching and angle bending are calculated with harmonic potentials, whilst the dihedral energies, $U(\varphi)$, are computed via Ryckaert-Bellemans functions, $U(\varphi) = \sum_{dihedrals} \sum_{n=0}^{5} C_n [\cos(\varphi - 180°)]$; we have included in the potential the refinements recently proposed by Lindorff-Larsen *et al.* [25], which result in improved accuracy of the DNA backbone dihedrals. The $Na^+$ and $Cl^-$ ions are described using the parameterization of Aqvist and Dang [26] and the water molecules by the TIP3P force field of Jorgensen and co-workers [27]; a recent discussion on the NaCl force-field influence upon the static and dynamic properties of nucleic acids under physiological conditions can be found elsewhere [26, 28]. To maintain computational tractability, we have chosen the double-stranded B–DNA Dickerson dodecamer [29, 30], exhibiting a pitch [31] of ~ 3.4 nm corresponding to an average of 10–10.5 base-pairs per turn over the entire helix [23], and with a double-strand end-to-end length of ~ 3.8 nm measured between terminal (GC) base pairs; the well-known A–DNA form has a pitch length of ~2.6 nm with an average of 11 base-pairs per turn [23]. Considering that the B–DNA backbone phosphorus atoms lie on a cylindrical surface, the average diameter of the double-strand corresponds to ~2 nm [31]. Although explicitly smaller in length than genomic DNA, the Dickerson dodecamer main structural features resemble those of genomic λ-bacteriophage DNA [32], namely in the radius of gyration and double-strand backbone diameter, $R_g \approx 0.7$–1 nm and $D \approx 2$ nm.

Recently, large diameter ($D \approx 4$ nm) single-walled carbon nanotubes have been prepared using supported-catalyst chemical vapour deposition [33]. In order to probe the confinement of DNA into such large, hollow nanostructures, we have adopted two different diameter SWCNTs with zig-zag symmetry, both with length $L = 8$ nm; the skeletal diameters, measured between carbon centres on opposite sides of the wall, are $D = 4$ nm (51,0) and $D = 3$ nm (40,0). The walls are built up of hexagonally-arranged $sp^2$ graphitic carbon atoms, with a C–C bond length [34, 35] of 1.42 Å, whose Lennard-Jones potential is given by Steele's parameterization ($\sigma = 0.34$ nm, $\varepsilon = 28$ K) [36]. The positions of all solid atoms are fixed throughout the calculations.

### B. Methodology and Algorithms

Large simulation cells with dimensions $(x, y, z) = 11 \cdot 11 \cdot 15$ nm were built according to the following stepwise procedure: *(i)* initially the solid was placed inside the empty cell, aligned along the $z$–axis, and the DNA molecule inserted at a distance of 0.5 nm away from the nanopore entrance, *(ii)* the whole cell was solvated with $H_2O$ ($\rho = 1$ g/cm$^3$) and the appropriate number of $Na^+$ and $Cl^-$ ions were added to ensure physiological ionic strength, [NaCl] = 134 mM, *(iii)* then the whole system was energy-minimized and equilibrated during at least 0.8 ns, first in the canonical ensemble and then in the isothermal-isobaric ensemble, resulting in a fully equilibrated simulation cell, as observed by the constancy of the main thermodynamical parameters (potential energy, temperature, volume and pressure). During minimization and equilibration the DNA position was constrained; once these steps were completed, the biomolecule was unconstrained, production runs were started and data collected over a time interval of 0.07–0.1 μs. The box dimensions were chosen in order to accommodate a 3.5 nm solvation shell in all directions around the nanotube.

Molecular dynamics (MD) simulations in the isothermal–isobaric ensemble ($NpT$) were performed using the Gromacs set of routines[37]. Newton's equations of motion were integrated with a time step of 1 fs and using a Nosé-Hoover thermostat [38, 39] and a Parrinello-Rahman barostat [40] to maintain temperature and pressure at 310 K and 1 bar. A potential cut-off of 1.5 nm was employed for both the van der Waals and Coulombic interactions, and the long-range electrostatics were calculated with the particle-mesh Ewald method [41, 42] using cubic interpolation and a maximum Fourier grid spacing of 0.12 nm. Three-dimensional periodic boundary conditions were applied. The well-tempered metadynamics scheme of Barducci and Parrinello [43] was employed to obtain the free-energy landscape associated with the confinement mechanism. The well-tempered algorithm biases Newton dynamics by adding a time-dependent Gaussian potential, $V(\zeta,t)$, to the total (unbiased) Hamiltonian, preventing the system from becoming permanently trapped in local energy minima and thus leading to a more efficient exploration of the phase space. The potential $V(\zeta,t)$ is a function of the so-called collective variables (or order parameters), $\zeta(q) = [\xi_1(q), \xi_2(q), ..., \xi_n(q)]$, which in turn are related to the microscopic coordinates of the real system, $q$, according to equation (1):

$$V(\xi(q),t) = W \sum_{t'=0}^{t'\leq t} \exp\left(-\frac{V(\xi(q(t')),t')}{\Delta T}\right) \exp\left(-\sum_{i=1}^{n}\frac{(\xi_i(q)-\xi_i(q(t')))^2}{2\sigma_i^2}\right) \quad (1)$$

where $t$ is the simulation time, $W=\tau_G\omega$ is the height of a single Gaussian, $\tau_G$ is the time interval at which the contribution for the bias potential, $V(\xi,t)$, is added, $\omega$ is the initial Gaussian height, $\Delta T$ is a parameter with dimensions of temperature, $\sigma_i$ is the Gaussian width and $n$ is the number of collective variables in the system; we have considered $\tau_G = 0.1$ ps, $\omega = 0.1$ kJ/mol, $\Delta T = 310$ K and $\sigma = 0.1$ nm. The parameter $\Delta T$ determines the rate of decay for the height of the added Gaussian potentials and when $\Delta T \rightarrow 0$ the well-tempered scheme approaches an unbiased simulation. The SWCNTs are primarily one-dimensional symmetric, therefore we decided to construct the free-energy landscape in terms of two collective variables, $\xi_1 = \vec{R}_z^{DNA} - \vec{R}_z^{SWCNT}$ and $\xi_2 = |\vec{R}_1^{GC} - \vec{R}_{12}^{GC}|$, where $\vec{R}$ is the positional vector of the centre of mass of the biomolecule ($\vec{R}_z^{DNA}$) and of the nanotube ($\vec{R}_z^{SWCNT}$), projected along the z–axis, or of the terminal (GC) nucleobase pairs at the double-strand *termini*, ($\vec{R}_1^{GC}$) and ($\vec{R}_{12}^{GC}$). According to our definition of collective variables, $\xi_1$ corresponds to the z-distance between the biomolecule and the nanopore centre and $\xi_2$ can be interpreted as the DNA end-to-end length. The characteristic lengths of the nanotube and Dickerson dodecamer are, respectively, $L = 8$ nm and $L = 3.8$ nm, and therefore any value $\xi_1 = \Delta L = (L^{SWCNT} - L^{DNA})/2 < 2.1$ nm corresponds to a DNA–SWCNT hybrid in which the biomolecule is completely encapsulated within the solid; the threshold $\xi_1 > 5.9$ nm obviously indicates the absence of confinement. At the end of the simulation, the three-dimensional free-energy surface is constructed by summing the accumulated time-dependent Gaussian potentials according to $F(\xi,t) = -\frac{T+\Delta T}{\Delta T}V(\xi,t)$. A discussion of the algorithm's convergence towards the correct free-energy profile is beyond the scope of this work and can be found elsewhere [43, 44]; suffices to say that it in the long time limit, $(\partial V(\xi,t)/\partial t) \rightarrow 0$, the well-tempered method leads to a converged free-energy surface. An alternative approach to obtain the time-independent free-energy surface relies on integrating $F(\xi,t)$ at the final portion of the metadynamics run [44]. The converged free-energy can thus be mathematically obtained from equation (2):

$$F(\xi) = -\frac{1}{\tau}\int_{t_{tot}-\tau}^{t_{tot}} V(\xi,t) \quad (2)$$

where $t_{tot}$ is the total simulation time and $\tau$ is the time window over which averaging is performed. We have implemented a convergence analysis for each collective variable, $\xi_1$ and $\xi_2$, splitting the last 40 ns of simulation time into $\tau = 4$ ns windows [30], and the results show that the bias potential $V(\xi, t)$ has converged, and thus the three-dimensional surface of Figure 2 is a good estimator of the free-energy changes associated with molecular encapsulation.

Independent calculations were performed using the umbrella sampling technique [45, 46]. For a system composed of $N$ particles, the method consists in biasing the classical (unbiased) Hamiltonian that depends on the potential, $U(r^N)$, and kinetic energies, $E_{kin}(p^N)$, by introducing a time independent harmonic potential, $V(\Omega_i) = \frac{1}{2}k(\Omega_i - \Omega_i^0)^2$, according to $H(r^N, p^N, \Omega_i) = U(r^N) + E_{kin}(p^N) + V(\Omega_i)$; $k$ is the harmonic force constant, $\Omega_i$ is an order parameter and $\Omega_i^0$ corresponds to the position of the umbrella restrain; in the present case, $\Omega_1 = |\vec{R}^{DNA} - \vec{R}^{SWCNT}|$ corresponds to the absolute three-dimensional distance between the centres of mass of the double-strand and the SWCNT, and $\Omega_2 = |\vec{R}^{r1r24} - \vec{R}^{r12r13}|$ is equivalent to the DNA end-to-end distance measured between *termini*. We have adopted $\Omega_1^0 = 0$ and $\Omega_2^0 = 4.1$ nm, in direct analogy with the collective variables defined in the well-tempered metadynamics algorithm, $\xi_1$ and $\xi_2$. When such a biasing potential is used, the biased probability distribution of the system, $P^b(\Omega_i)$, can be obtained from a Boltzmann weighted average along the $\Omega_i$ order parameter and, therefore, assuming that the system is ergodic [46]:

$$P^b(\Omega_i) = \frac{\int \exp\{-\beta[U(r)+V(\Omega_i'(r))]\}\delta[\Omega_i'(r)-\Omega_i]d^Nr}{\int \exp\{-\beta[U(r)+V(\Omega_i'(r))]\}d^Nr} \quad (3)$$

where $\beta = (1/k_BT)$, $k_B$ is the Boltzmann constant, $\delta$ is the Dirac delta function, and $N$ is the total number of particles in the system. Because the biasing potential depends only on the order parameter $\Omega_i$, and the integration in the numerator is performed over all degrees of freedom except $\Omega$, the unbiased probability of the real system, $P^u(\Omega_i)$, can be evaluated from equation (4):

$$P^u(\Omega_i) = P^b(\Omega_i)\exp^{\beta V(\Omega_i)}\Gamma \quad (4)$$

where $\Gamma = -(1/\beta)\ln\langle e^{-\beta V(\Omega)} \rangle$ is independent of $\Omega_i$ and the triangular brackets denote an ensemble average. The reconstruction of the true (unbiased) free energy profile or potential of mean force [47], consistent with the Gibbs free energy, $PMF(\Omega) = -k_B T \ln P^u(\Omega)$, is accomplished using the weighted histogram analysis method [47-49].

### C. Parametric Analysis

The radius of gyration, $R_g$, gives a measure of a molecule's compactness and is defined by equation 5:

$$R_g = \sqrt{\left(\sum_i^N \|r_i\|^2 m_i\right) \bigg/ \sum_i^N m_i} \quad (5)$$

where $N = 758$ is the total number of atoms in the DNA molecule, $m_i$ is the mass of atom $i$ and $r_i$ is the positional vector of the atom relative to the molecular center of mass. The root mean squared deviation, $RMSD$, is obtained by calculating the distance $r_{ij}$ between atoms $i$ and $j$ at time $t$, and comparing with the same distance observed at time $t = 0$, according to:

$$RMSD = \sqrt{(1/N^2)\sum_{i=1}^N \sum_{j=1}^N \|r_{ij}(t) - r_{ij}(0)\|^2} \quad (6)$$

## III. RESULTS AND DISCUSSION

Confinement of double-stranded DNA (dsDNA, [30]) into a (51,0) nanotube ($D = 4$ nm) is fast and becomes complete within the first 16 ns of observation time. Initially, the double strand is in the bulk (0–2 ns) and as it diffuses towards the SWCNT entrance undergoes structural rearrangements leading to minor increases in pitch length, $P$, and end-to-end distance, $L$ (Fig. 1A). After 2 ns, the dodecamer is already at the nanopore entrance, where it experiences strong van der Waals attractions towards the solid [30], resulting in complete encapsulation at 15.4 ns, after which the double-strand relaxes to $P \approx 3.4$ nm and $L \approx 4 - 4.1$ nm. It is very interesting to observe that confinement appears to be permanent, *i.e.*, the DNA fragment never returns to the bulk solution during the observation time window, maintaining direct local contact with the solid wall at a distance of closest approach of *ca.* 2.6 Å. Nonetheless, the encapsulated molecule clearly retains its translational mobility, diffusing freely along the nanopore main axis (Fig. 1B). The encapsulation mechanism can be divided into a three-step process: *(i)* fast diffusion of DNA towards the nanopore entrance (0–2 ns), *(ii)* strong van der Waals attractions towards the solid, leading to confinement of *terminus* 1 at 2.33 ns, followed by structural rearrangements of the whole double strand occurring in bulk solution ($t < 15$ ns) and finally *(iii)* penetration of *terminus* 2 into the confining volume resulting in complete encapsulation of the biomolecule.

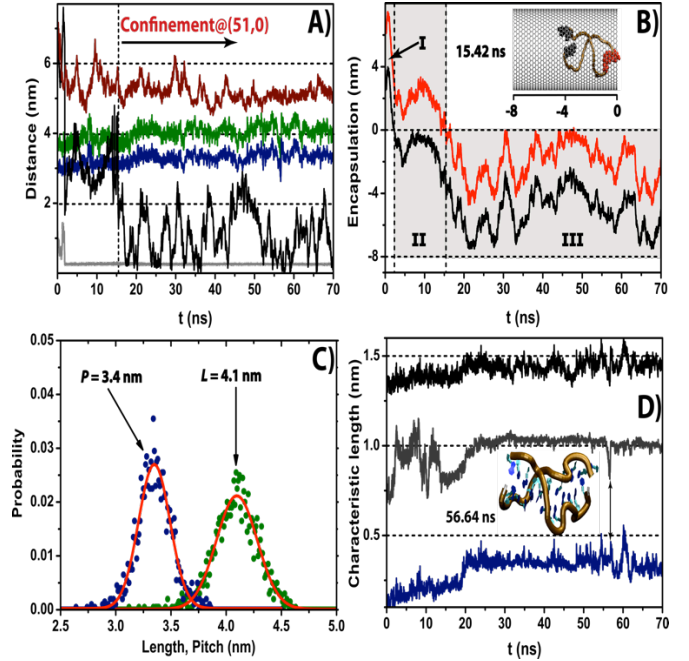

**Figure 1. Encapsulation of double-stranded DNA into a (51,0) SWCNT at $T = 310$ K and [NaCl] = 134 mM.** *A) Kinetics.* Encapsulation is complete after 15.4 ns after which the double strand never returns to the bulk solution during the observation time window; nonetheless, DNA maintains its translational mobility within the nanopore: (black) distance between centres of mass (c.o.m.s) of DNA and the (51,0) SWCNT, (dark red) distance between c.o.m.s of DNA and the (40,0) SWCNT, (green) distance between opposite GC *termini* in the double strand, *e.g.*, DNA end-to-end length, $L$, (blue) DNA pitch length, (grey) minimum distance between any DNA atom and the SWCNT. A 2 nm c.o.m. distance between the biomolecule and the solid corresponds to the threshold bellow which complete encapsulation is considered to occur. *B) State diagram of the encapsulation mechanism.* Lines represent distances between terminal nucleobase pairs and the SWCNT entrance through which encapsulation takes place, projected along the nanotube main axis: (black) terminus 1 and (red) terminus 2. The horizontal dashed lines correspond to the SWCNT boundaries and the inset magnification depicts the first instant just after complete encapsulation occurs, representing the terminus 1 atoms in black and the terminus 2 atoms in red. Note that the confined molecule maintains its translational mobility along the nanopore's axial axis. *C) Probability distribution profiles of DNA encapsulated into a (51,0) SWCNT.* Pitch (blue dots) and DNA length (green dots); red lines correspond to Gaussian fittings of simulation data. *D) DNA characteristic lengths.* (black) radius of gyration, $R_g$, (grey) z-component of the radius of gyration, $R_g^z$, (blue) root mean-squared deviation, $RMSD$. After confinement, molecular conformation deviates minimally from the canonical B-DNA form, $RMSD^{aver} = 0.36 \pm 0.002$ nm, and maintains a quasi-linearity about the nanopore main axis.

In contrast, the narrowness of a (40,0) topology ($D = 3$ nm) completely inhibits encapsulation even over an observation time of 0.1 μs. Instead, the DNA contact with the smaller nanopore results in the occurrence of two distinct states: *(i)* exoadsorption onto the external surface of the nanotube via a π–π stacking mechanism of a terminal (GC) nucleobase pair onto the graphitic surface, and threading of the biomolecule along the hydrophobic cylinder similar to previous observations [11], or *(ii)* trapping of the DNA at the nanopore entrance, with partial melting of the double-strand

*terminus* closest to the solid; Mogurampelly *et al.* [22] observed that encapsulation onto a (20, 20) tube ($D$ = 2.67 nm) is thermodynamically prohibited due to a large free-energy barrier located at the nanopore entrance. The decomposition of the interaction energies between DNA and the surrounding environment, solid and solution [30], can help to throw some light onto this issue. Upon confinement onto the (51,0) topology, the dodecamer becomes less solvated by the $H_2O$ molecules than in the bulk, as indicated by a decrease of the DNA/$H_2O$ interaction energy of 1327 kJ/mol, and leading to a thin hydration shell of *ca.* **1 – 2** water molecules between the biomolecule and the solid walls. However, this instabilization is roughly compensated by an increase of the DNA/ions interactions of – 851 kJ/mol and also by the intrinsic effect exerted by the hydrophobic solid upon the DNA van der Waals cloud of – 442 kJ/mol. On the other hand, molecular exoadsorption onto the (40,0) nanotube prevents stabilization of the biomolecule from the dispersive interaction with the graphitic walls, leading to a diminished DNA/(40,0) interaction energy of – 170 kJ/mol, clearly insufficient to overcome the energetic penalty of a decreased solvation effect caused by encapsulation.

At physiological conditions the canonical B-form is dsDNA most stable configuration [31, 50]; however, little is known when the molecule is confined into a strongly hydrophobic solid, such as a carbon nanotube. The data recorded in Fig. 1A suggest that this is also the case for encapsulated dsDNA. In fact, when simulation results are used to determine distribution histograms, the corresponding frequencies are well correlated by Gaussian statistics exhibiting a pitch length of $P$ = 3.4 nm, for a 10 nucleobase pair repeating unit, and a double strand length of $L$ = 4.1 nm (Fig. 1C), consistent with the geometrical characteristics of B-DNA in bulk solution [23]. We have determined the biomolecular characteristic lengths—radius of gyration ($R_g$, equation 5) and its projection along the nanotube main axis ($R_g^z$) as well as the root-mean squared deviation (RMSD, equation 6) from the B-DNA Dickerson dodecamer used as starting configuration [30]—and plotted them in Fig. 1D. Because the RMSD compares the structure at any time $t$ with the original DNA structure ($t$ = 0), the blue curve in Fig. 1D indicates a minor relaxation of the double-strand from the crystal structure to accommodate liquid state flexibility, whilst maintaining the relative average distance between each atom in the double-strand. After 20 ns, the data converged smoothly to average values of RMSD = 0.36 ± 0.002 nm and $R_g^z$ = 1.02 ± 0.0015 nm; the latter value indicates an alignment of the biomolecule along the axial axis of the nanopore. The $R_g^z$ = 0.8 nm depression observed at 56.6 ns is transient and matches a total number of Watson-Crick H bonds of 28.

The thermodynamical stability of molecular encapsulation is probed by the free-energy ($F$) differences associated with the process, using well-tempered metadynamics [43, 44]. We have chosen two order parameters to construct the free-energy landscape, $\xi_1$ and $\xi_2$, where $\xi_1$ is the distance between centres of mass of DNA and SWCNT, projected along the nanopore main axis ($z$), and $\xi_2$ is the absolute distance between the (GC) *termini* on opposite sides of the double-strand, equivalent to the DNA end-to-end length (*cf.* Models and Methods). An inspection of the resulting 3–D surface (Figure 2) reveals the existence of five distinct free-energy minima, sharing in common the fact that all are located at discrete positions along the internal volume of the nanopore, $\xi_1$ < 1.8 nm; the absolute minimum at $\xi_1$ = 0.117 nm indicates that the center of the nanotube is the most energetically stable region for the encapsulated biomolecule, which results in the strongest concentration of molecular density at that location [30]. Owing to their thermodynamical similarity, the five free-energy minima along $\xi_1$ provide a minimum free-energy path along which DNA can translate within the pore, visiting maximum probability configurations as indicated by the dotted line in Figure 2. In order to escape from those deep free-energy valleys, $F$ ($\xi_1$, $\xi_2$) ~ – 40 kJ/mol$^{-1}$, DNA has to overcome large energetic barriers, rendering the exit process towards the bulk solution thermodynamically expensive. The reversibility of encapsulation is discussed in the Conclusions, where different ejection mechanisms are tackled to externalize the biomolecule.

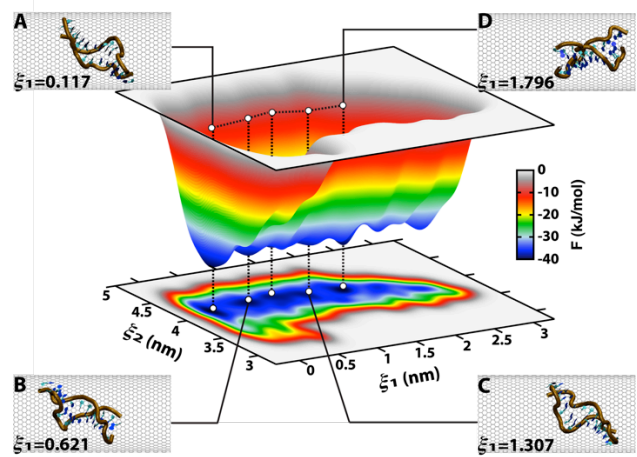

**Figure 2. Free energy landscape of DNA@(51,0) SWCNT hybrid.** The thermodynamical surface is built using two macroscopic descriptors, $\xi_1$ and $\xi_2$, where $\xi_1$ is the distance between centres of mass of the dsDNA and nanotube, projected along the nanopore main axis ($z$), and $\xi_2$ is the absolute distance between the (GC) *termini* on opposite sides of the double-strand, equivalent to the dsDNA end-to-end length. The several adjacent free-energy minima along $\xi_1$ demonstrate that the molecule is relatively mobile to translocate along the nanotube, however, the absolute minimum at $\xi_1$ = 0.117 nm indicates the nanopore center is the energetically favoured region upon confinement. All the $\xi_1$ minima are located along a *quasi*-linear path defined by $\xi_2$ ≈ 4.1 nm highlighting the enhanced thermodynamic stability associated with the canonical B form. The snapshots were taken at different time intervals corresponding to ($\xi_1$, $\xi_2$) nm: *A)* (0.117, 4.112), *B)* (0.621, 4.164), *C)* (1.307, 4.164) and *D)* (1.796, 4.115). $H_2O$ molecules and $Na^+$ and $Cl^-$ ions are omitted for clarity sake.

It is remarkable to observe that the exact position of the free-energy minima is almost invariant at $\xi_2$ ~ 4.1 nm, consistent with the end-to-end length of $L$ = 3.8 nm of a B–

DNA conformation. We have performed independent umbrella sampling calculations, using an harmonic bias, to determine the potential of mean force (PMF) and the system's thermodynamical probability distribution, $P(\Omega_i)$, employing two order parameters to describe the distance between c.o.m.s of DNA and the solid, $\Omega_1$, and the end-to-end length of the double-strand, $\Omega_2$. The calculations with $\Omega_1$ clearly corroborate the main findings revealed by the metadynamics analysis, namely that the system has probability maxima at $\Omega_1$ = (0.21, 1.62, 1.92) nm (Figure 3); the DNA molecule is fairly mobile inside the nanopore, easily moving from one free-energy minimum to an adjacent one. The thermodynamical robustness of the canonical B-form under confinement is illustrated by the Gaussian profile assumed by the probability distribution regarding the DNA end-to-end length, Figure 3. $P(\Omega_2)$ is well described by

$$P(\Omega_2) = \varphi \exp\left[-\frac{1}{2}(\Omega_2 - \Omega_2^0/\sigma)^2\right],$$

with a Gaussian peak width at half height of $\sigma = 0.16$ nm, $\phi = 4.32 \cdot 10^{-2}$, and a peak centered at $\Omega_2^0 = 4.01$ nm ± 0.001 nm corresponding to the equilibrium (unbiased) end-to-end distance of encapsulated dsDNA, $L$. It now becomes clear that a perturbation of the double strand towards non-equilibrium values of $L$, leading to either a contraction $\Omega_2 < 4.01$) or a stretching $\Omega_2 > 4.01$) of B-DNA, results in a rapid increase of the associated potential of mean force rendering the perturbation process thermodynamically unstable.

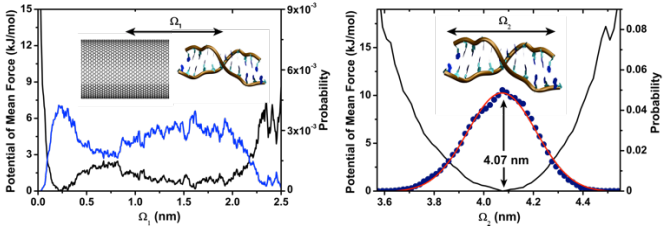

**Figure 3. Potential of mean force, PMF (black), and probability distribution (blue line, blue dots).** $\Omega_1$ is the order parameter defined by the distance between centres of mass of DNA and the nanotube and $\Omega_2$ is the end-to-end distance of the double-strand. The $\Omega_2$ probability distribution curve is Gaussian shaped and the red line is the best fit of $P(\Omega_2) = \varphi \exp\left[-\frac{1}{2}(\Omega_2 - \Omega_2^0/\sigma)^2\right]$ to the umbrella data.

In spite of the thermodynamical stability of the encapsulated B-form, the geometric characteristics of the double strand naturally oscillate about their equilibrium values, such as the angles between each individual strand *termini*, $\varphi_{ij}$. Postulating that $C$ corresponds to the DNA molecular center and is given by the average axes bridging the individual strands (Figure 4), then

$$\phi_{ij} = \angle(\vec{r}_i \cdot C \cdot \vec{r}_j) = \arccos\left[\left(\|\vec{r}_i\|^2 + \|\vec{r}_j\|^2 - \|\vec{r}_i - \vec{r}_j\|^2\right)/\left(2\|\vec{r}_i\|\|\vec{r}_j\|\right)\right]$$

, where $\vec{r}_i$ is the positional vector, with origin at $C$, of a terminal phosphorus atom belonging to nucleotide $i$. To simplify the notation, let $\phi_1 = \phi_{224} = \angle(r_2 \cdot C \cdot r_{24})$ and $\phi_2 = \phi_{1214} = \angle(r_{12} \cdot C \cdot r_{14})$, which for the crystalline form of pure B-DNA assume the values $\varphi_1 = 55.9°$ and $\varphi_2 = 56.6°$ as measured by Dickerson and co-workers [29]. The anisotropy of the terminal angles is simply given by $\Delta\varphi = (\phi_1 - \phi_2)$. The local elevation of $\Delta\varphi$ observed in the 10–20 ns time window, an interval during which DNA is being confined, falls back to negligible values ($\Delta\varphi \approx 0.35°$) indicating that the anisotropic deformation of the double strand is reversible once confinement is complete. Because the molecule undergoes encapsulation from the $\varphi_2$ *terminii*, the fact that $\varphi_1 > \varphi_2$ while confinement takes place indicates a slight compression at the $\varphi_2$-end, which is replicated on the other side of the chain ($\varphi_1$) as it penetrates into the SWCNT. This previously unobserved compression phenomenon is an entropic effect arising from the constriction caused by the pore walls, but also an energetic effect given that the interaction energy between DNA and the nanotube only stabilizes once encapsulation is complete [30]. To accommodate itself in the endohedral volume, the double-strand skeletal diameter slightly decreases and is balanced by an end-to-end length increase from $L = 3.85$ nm to $L = 4$–4.1nm (Figure 1A).

Upon encapsulation the dodecamer maintains its kinetic mobility, exploring a region whose boundaries are located at the nanotube *termini* and correspond to minima in the overall free-energy landscape (Figure 2). Translocation within the solid occurs via a self-translational diffusion process along the central axis and also via a mechanism of self-rotation about the biomolecular axis. The conformational *ensemble* of individual strand axes recorded in Figure 4 resembles a toroid in the $yz$-plane (parallel to the nanotube main axis), whose centre is largely unpopulated. On the other hand, even though the double strand is flexible, it cannot be overstretched along the $z$-axis without a drastic increase of the corresponding PMF, and thus the $yx$-projection shows symmetrical opposite regions at the boundaries (corresponding to domains close to the wall), where the density of the axes is smaller.

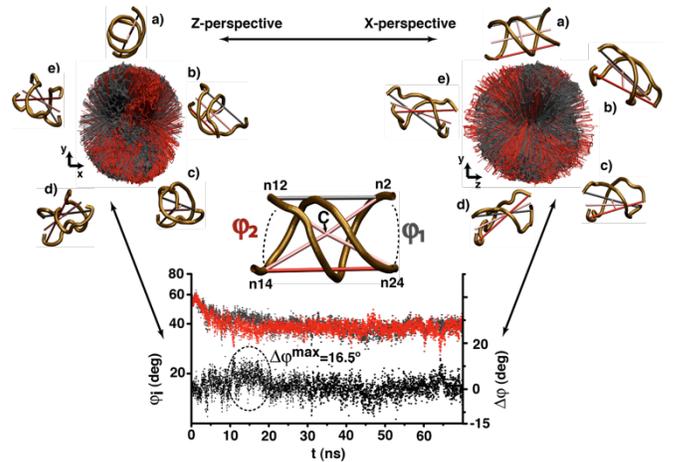

**Figure 4. Top:** *Quasi*-isotropic individual axis distribution for confined DNA. (grey) axis corresponding to strand A, running from nucleotide 2 up to nucleotide 12, (red) axis corresponding to strand B ($n14$-$n24$). The double-helix representation is as follows: (ochre), DNA backbone, (grey and red), single-strand axis, (pink) double-strand average axes; (*a*) $(\varphi_1, \varphi_2) = (55.9, 56.6)°$, crystalline B-DNA Dickerson structure, $t = 0$ ns; (*b*) $(\varphi_1, \varphi_2) = (48.36, 31.84)°$, $t = 12.47$ ns; (*c*) $(\varphi_1, \varphi_2) = (41.5, 37.9)°$ $t = 17.5$ ns (during confinement); (*d*) $(\varphi_1, \varphi_2) = (40.7, 37.4)°$, $t = 30$ ns; (*e*) $(\varphi_1, \varphi_2) = (39.8, 38.5)°$, $t = 70$ ns. Notice the almost parallel alignment of the individual strand axis observed in the crystalline Dickerson dodecamer. H$_2$O molecules and NaCl ions are omitted for clarity. **Bottom: Temporal evolution of the inter-strand terminal angles.** The double-strand backbone is coloured ochre and the single-strand individual axes are coloured grey and red. The log-scale on the left reads the individual double-strand terminal angles, $\varphi_1$ (grey) and $\varphi_2$ (red), whilst the linear right-hand-side scale indicates the nominal difference between both angles, $\Delta\varphi = (\varphi_1 - \varphi_2)$ (black).

The internal structure of the double-helix, $\Xi$, is probed by measuring the minimum distance of closest approach between each nucleotide, $d(\Psi_i, \Psi_j)_{i,j=1-24}$. The resulting contact maps in Figure 5 indicate that $\Xi$ is essentially maintained invariant, prior to, during, and after DNA encapsulation. It should be noted that the contact map recorded at $t = 0$ ns corresponds to the pure crystalline form of the B-DNA Dickerson dodecamer. Because the distance between a nucleotide and itself is null, the dark blue diagonal is related with the intrastrand structure; adjacent nucleotides that belong to the same strand are always at a distance $d(\Psi_i, \Psi_j)_{j=i+1} < 0.4$ nm. The light blue diagonal represents the stability of the interstrand structure, determined by H-bonding between complementary nucleotides located in different strands. At 15.42 ns there is a slight increase in contact distance around an area defined by the terminal pair (12-13), in direct correspondence with the anisotropic deformation highlighted in Figure 4. This lateral opening of the chain, resulting in an encapsulation anisotropy of $\Delta\varphi^{max}=16.5°$, is clearly a reversible process because $d(\Psi_{12}, \Psi_{13})$ falls back to the purely crystalline B-DNA values during the observed time window. The two symmetrical shoulders located at regions defined by nucleotide indices (20-22) and (8-10), evidencing a slight distance decrease between contacts, need to be carefully analysed.

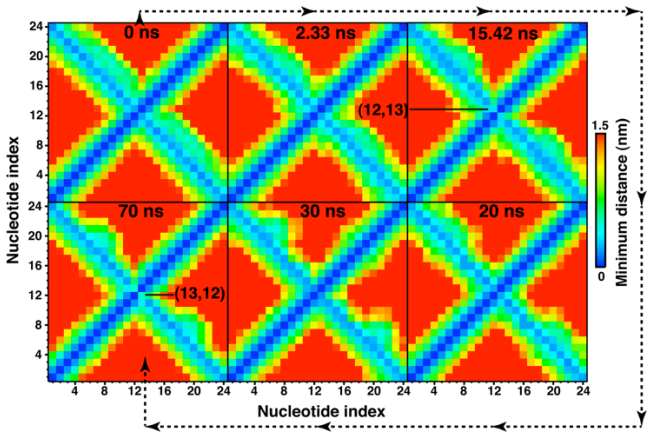

**Figure 5. Contact maps between the DNA nucleotides.** The distance between two nucleotides $\Psi_i$ and $\Psi_j$ is defined as the minimum distance between any pair of atoms ($i, j$: $i \in \Psi_i, j \in \Psi_j$); by definition $d(\Psi_i, \Psi_j) = d(\Psi_j, \Psi_i)$. The terminal nucleotide pairs, defining the DNA persistence length, are the first nucleotide H-bonded to the 24[th] nucleotide and nucleotide 12 H-bonded to the 13[th] one. Strand A is formed from nucleotide 1 to 12 and the (complementary) B strand runs from nucleotides 13 to 24. Adjacent nucleotides that belong to the same strand are always at a distance $d(\Psi_i, \Psi_j) < 0.4$. Notice the slight increase in contact distance around the terminal pair (12-13).

The conformational stability of DNA is markedly influenced by the canonical Watson-Crick H-bonds network, which is also paramount to maintain double-strand integrity and avoiding melting into single strands. In crystals, the distance between a donor (D) and the acceptor (A), $d_{D-A}$, is maximum for the N6–O4 H-bond of (AT) pairs corresponding to $d_{D-A} = 2.95$ Å (Figure 6A), however, in the liquid phase $d_{D-A}$ slightly increases to accommodate transient thermal fluctuations and molecular flexibility. Accordingly with previous studies [18, 19, 31] we postulate that an H-bond exists when the donor and acceptor are not separated by more than $d_{D-A} = 3.5$ Å and with a characteristic angle [19] of $\angle_{D-H-A} \leq 35°$. In order to probe the complete Watson-Crick network, we have determined the number of H-bonds of each donor-acceptor type occurring in the DNA dodecamer and represented them in terms of the corresponding probability of occurrence computed by histogram weighting. The results in Figure 6A are represented in terms of the normalized probability of occurrence, $P(\%)$, where the normalization $\int_0^{100\%} P(\%)d\% = 1$ is performed for the maximum number of allowed H-bonds for a particular donor-acceptor pair. The H-bond between N1 and N3, characteristic of both (AC) and (GC) basepairs, has a maximum occurrence allowance of 12, and therefore the peak at 92% with a normalized probability of $P(92\%) = 0.574$ corresponds to the existence of 11 H-bonds; the other occurrences are located at 100% with $P(100\%) = 0.095$ (12 H-bonds), at 83.5% with $P(83.5\%) = 0.264$ (10 H-bonds), at 75% with $P(75\%) = 0.061$ (9 H-bonds) and at 67% with $P(67\%) = 0.006$ (8 H-bonds). Occurrences bellow 50% have negligible probabilities for all H-bonds in the double-strand. For occurrences higher than 75%, the accumulated probabilities correspond to $P(\% > 75) \geq 0.87$–$0.91$; thus, Figure 6A indicates that the canonical Watson-Crick H-bond network is essentially conserved when the molecule becomes encapsulated under physiological conditions. This comes to show that the contact between the DNA dodecamer and the hydrophobic inner surface of the carbon nanotube results in rather minor rearrangements of the nucleotides H-bonding, which is of the utmost relevance to maintain double-strand integrity for drug delivery techniques currently exploring SWCNT-based media as encapsulating agents [6, 7].

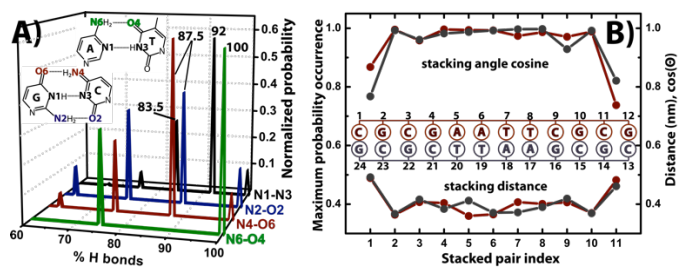

**Figure 6. A) Canonical Watson-Crick H-bonds.** Normalized probability distributions, $\int_{0}^{100\%} P(\%)d\% = 1$, of the percentage of canonical H-bonds present in the encapsulated DNA molecule, considering the last 20 ns of observation time: for the N1–N3 pair, %H bonds = 100% corresponds to the existence of 12 H-bonds throughout the dodecamer, %H bonds = 92% to 11 H-bonds, %H bonds = 83.5% to 10 H-bonds and so on. The probability maxima are indicated in the graph by the corresponding %-value: (dark blue) H-bonds between N2–O2, (red) H-bonds between N4–O6, (green) H-bonds between N6–O4 and (black) H-bonds between N1–N3. **B) Intrastrand stacking geometry.** (Dark red) strand 1, (grey) strand 2. The stacking distance (helical rise) is measured between the geometrical centres of the molecular planes [30] belonging to two consecutive nucleobases of the same strand (*e.g.*, stacked NB pair 1 corresponds to C1–G2 and also G24–C23), and the stacking angle is calculated between the normal vectors of those two planes (two perfectly stacked bases, with their rings parallel to each other, have a stacking angle of 0° corresponding to cos (0°) = 1).

The N2–O2 bond in (GC) pairs shows an accumulated probability of $P(\%>75) \sim 0.87$, suggesting that H-bond breaking/formation occurs essentially in the double-strand *termini*, richer in (GC) moieties, leaving intact the (AT) rich inner tract. This is clearly consistent with the intrastrand data recorded in Figure 6B, where consecutive nucleotides have been probed for their stacking distance (helical rise) and normal vector angle between molecular planes. From the corresponding distribution histograms, the maximum probability of occurrence has been obtained for both geometric parameters and plotted as function of the stacked pair index. The crystallographic studies of Vargason *et al.* [23] indicate that the transition from pure B-DNA to an A-DNA conformation is accomplished via a monotonical decrease of the intrastrand stacking distance from 3.4 Å to *ca.* 2.6 Å for the pure A-form. Apart from the severe deformation induced by basepairs 1 and 11 upon the double-strand, our results indicate an helical rise between nucleotides of ~3.8 Å, resulting in $L = 4.18$ nm. We have observed (Figure 1A) that encapsulation of DNA leads to a slight increase of $L$ from the pure B-form, and are now able to attribute this increase to a 1D anisotropic stretching of the molecule located essentially at the end of the (GC) tracts. From an energetical point of view this is somehow unexpected, bearing in mind that (AT) pairs form two H-bonds instead of the characteristic three H-bonds of (GC) duos; the interaction energies between purines and pyrimidines favour the stability of (GC) pairs against their (AT) counterparts, $E_{GC} = –291.9$ kJ/mol and $E_{AT} = –110.2$ kJ/mol [30]. The explanation for this apparent inconsistency lies in entropic causes. Because the *termini* are more flexible than the inner core, the former are more prone to adaption to the local environment, thus being more able to accommodate elastic deformation.

## IV. CONCLUSIONS AND PROSPECTS

DNA can be encapsulated onto the purely hydrophobic (51,0) topology but not so at the (40,0) analogue, indicating that a 1 nm decrease of diameter might prevent confinement. This observation is encouraging for technologies using pristine SWCNTs as drug delivery agents, however, needs to be wisely put in perspective. It is known that nanotubes can be electrically charged, either using an AFM tip and applying a voltage bias or by chemically doping the solids with p-type dopants to obtain positively charged nanotubes [11, 51]. The effect of charge density upon the energetics and dynamics of confinement needs to be carefully addressed in the future; because DNA's outer surface is negatively charged (phosphates), its interaction with a positively charged (40,0) solid might indeed lead to the occurrence of encapsulation with enhanced thermodynamical stability. The latter is of paramount importance, for any technological application to find its way into the industrial production line the confinement of DNA molecules needs to be a thermodynamically reversible process, and subsequent ejection possible towards the nanotube exterior. Xue et al used filler agents ($C_{60}$) and mechanical actuators (Ag) to eject ssDNA from hydrophobic SWCNTs [52], and related the feasibility of the ejection process with the enhanced dispersive interactions resulting from DNA externalization; their $C_{60}$ agents evidenced interaction energies with the nanotube of ~ –1800 kJ/mol (~ –7000 kJ/mol for the Ag nanowires), an order of magnitude higher than the ones we observe for the (51,0) encapsulated DNA [30], leading us to believe that similar externalization mechanisms can be employed to revert DNA encapsulation.

### ACKNOWLEDGEMENTS

The authors would like to acknowledge Requimte/CQFB (Universidade Nova de Lisboa, Portugal) and NSEC (University of Wisconsin – Madison, USA) for generous CPU time. This work was partially supported by grant PTDC/CTM/104782/2008 (Portugal) and makes use of results produced with the support of the Portuguese National Grid Initiative (https://wiki.ncg.ingrid.pt). F.J.A.L. Cruz gratefully acknowledges financial support from FCT/MCTES (Portugal) through grant SFRH/BPD/45064/2008.